\documentstyle [12pt] {article}

\catcode`@=12
\begin {document}
\thispagestyle{empty}
\begin{flushright} {UCRHEP-T93\\June 1992}
\end{flushright}
\begin{center}
\vspace{0.5in}
{\Large \bf Accidental Approximate Generation Universality\\}
\vspace{0.3in}
{\Large \bf and its Possible Verification$^*$\\}
\vspace{1.2in}
{\bf Ernest Ma\\}
\vspace{0.3in}
{\sl Department of Physics\\}
{\sl University of California\\}
{\sl Riverside, California 92521\\}
\vspace{1.2in}
\end{center}
\begin{abstract}\
The universality of $e-\mu-\tau$ interactions may only
be an accidental approximate symmetry analogous to that of flavor $SU(2)$ and
$SU(3)$.  This was specifically realized by an extension of the standard model
proposed in 1981.  Two key predictions are that the $\tau$ lifetime should be
longer and that the $\rho$ parameter measured at the $Z$ peak should have an
additional negative contribution.  These are consistent with present precision
electroweak measurements.   A future decisive test of this model would be the
discovery of new $W$ and $Z$ bosons with nearly degenerate masses of a few
$TeV$.
\end{abstract}\
$^*$To appear in {\em Proc. of Beyond the Standard Model III}.
\newpage
\section{Introduction}

	Historically, flavor $SU(2)$ and $SU(3)$ were thought to be fundamental,
albeit approximate, symmetries.  We now know that the real fundamental symmetry
is color $SU(3)$, and that the former are merely accidental symmetries:  flavor
$SU(2)$ is the result of the hierarchy $m_u, m_d << \Lambda_{QCD}$, and flavor
$SU(3)$ is even less exact because $m_s$ is not as negligible.

	Consider now generation universality.  In the standard model, the
implicit assumption is that it is a fundamental symmetry resulting from the
existence of 3 (or more) generations of quarks and leptons which have identical
gauge interactions. On the other hand, this may be the wrong way of looking at
it.  Generations could be fundamentally different and their interactions at
low energies may be only accidentally and approximately universal, in analogy
with flavor $SU(2)$ and $SU(3)$.  This alternative viewpoint was
specifically realized in a model by X. Li and myself, published in
1981.\cite{1}

	The 2 main predictions of this model relevant to present electroweak
measurements are: (1) the $\tau$ lifetime should be longer than predicted by
the standard model; and (2) the $\rho$ parameter measured in the decay of $Z$
into leptons should have an additional negative contribution.  Both trends are
now noticeable in the data, although one certainly cannot claim that
definite deviations from the standard model have been observed.

\section{The Model}

	Consider the gauge group $U(1)~X~SU(2)_1~X~SU(2)_2~X~SU(2)_3$ with
couplings $g_0,g_1,g_2,g_3,$ respectively.  The left-handed fermions are
doublets under $U(1)~X~SU(2)_i$, with each generation coupling to a separate
$SU(2)$.  The right-handed fermions are singlets coupling only to $U(1)$.
This structure is anomaly-free and has no redundancies.  The Higgs sector
consists of 3 $SU(2)_i$ doublets with vacuum expectation values $v_{0i}$ and
3 $SU(2)_j X SU(2)_k$ self-dual bidoublets with vacuum expectation values
$v_{jk}$.  This then guarantees the equality of the effective weak-interaction
coupling matrix $(G_F)_{ij}$ for charged and neutral currents, as in the
standard model.  If $v_{01}^2+v_{02}^2+v_{03}^2 << v_{12}^2$, then
$(G_F)_{ij} = G_F$ for $i\neq 3$ and $j\neq 3$, and $(G_F)_{ij} = \xi ^{-1}
G_F$ for $i=3$ or $j=3$ but not both, where $\xi \equiv 1+v_{03}^2/(v_{13}^2+
v_{23}^2)$.  Hence $e-\mu$ universality is valid, but $e-\mu-\tau$ universality
is only approximate. In particular, the $\tau$ lifetime should be longer than
the standard-model prediction by the factor $\xi^2$.  Experimentally, from
all the data available up to several months ago, $\xi-1$ was determined to be
$0.027\pm0.012$,\cite{2} which was a 2.3$\sigma$ effect.  However, mainly
because of the new measurement of $m_{\tau} = 1776.9 \pm 0.4 \pm 0.3~MeV$
by the BES collaboration at the BEPC $e^+e^-$ collider in Beijing, it has now
been reduced to $0.015\pm0.008$, which is only a 1.8$\sigma$ effect in support
of our prediction.

	Let us also define $r \equiv (v_{01}^2+v_{02}^2)/v_{03}^2$ and
$y \equiv g_{123}^2/g_3^2$, where $g_{123}^{-2} = g_1^{-2}+g_2^{-2}+g_3^{-2}$.
Then $e^{-2}~=~g_0^{-2}+g_{123}^{-2}$, and for the first 2 generations,
\begin{equation}
H_{NC}^{eff}~=~{4G_F \over \sqrt 2} \left[ (j^{(3)} - s^2 j^{em})^2~+~
C(j^{em})^2 \right],
\end{equation}
where
\begin{equation}
s^2 = 1-e^2/g_0^2-(1-\xi^{-1})e^2/g_3^2,
\end{equation}
and
\begin{equation}
C = (e^4/g_3^4)(1-\xi^{-1})(\xi^{-1}+r)\simeq(\xi-1)~s^4y^2(1+r).
\end{equation}
The above effective interaction is
the result of the virtual exchange of all 3 $Z$ bosons of this model, but
at LEP, only one of them (the lightest) is produced and it should certainly
not be identical to the standard-model $Z$.  In fact, its mass squared is
given by $1+(\xi-1)y((1-s^2)^{-1}-(1+r)y)$ times that of the standard model,
as predicted from the value of $s^2$ in Eq. 1.  The corresponding factor for
$M_W^2$ is $1+(\xi-1)y(1-(1+r)y)$.

\section{Z Leptonic Widths}

	From precision measurements of the widths and forward-backward
asymmetries of $Z \rightarrow l\overline{l}~(l=e,\mu,\tau)$ at LEP, the
parameters $\rho_l$ and $sin^2\theta_l$ are extracted.
\begin{equation}
\Gamma_l~=~{{G_F M_Z^3} \over {24 \sqrt 2 \pi}} \left( 1+{{3 \alpha} \over {4
\pi}} \right) \rho_l \left[ 1+\left( 1-4sin^2\theta_l \right)^2 \right],
\end{equation}
and
\begin{equation}
A_{FB}^l (M_Z^2)~\simeq~3 \left( 1-4sin^2\theta_l \right)^2.
\end{equation}
In the nonuniversality model $\rho_{e,\mu}=1-(\xi-1)y^2(1+r)+\rho_{rad}$,
where $\rho_{rad}$ is dominated by the standard-model $m_t^2$ contribution.
[The Higgs sector of this model has an automatic $SU(2)$ custodial symmetry
which eliminates all the quadratic mass terms of the scalar bosons to one-loop
order.]  Since $\xi>1$, $r>0$, and $0<y<1$ are required by their definitions,
this model has a necessarily negative contribution to $\rho_{e,\mu}$, in
addition to the necessarily positive contribution of $\rho_{rad} \simeq
3\sqrt2G_Fm_t^2/16\pi^2$ of the standard model.

	Let $x \equiv y(1+r)$, then $\rho_\tau = \rho_{e,\mu} - 2(\xi-1)(1-x)$
and
\begin{equation}
{\Gamma_\tau \over \Gamma_{e,\mu}}~\simeq~1-{{2(\xi-1)(1-2s^2)} \over {1-4s^2+
8s^4}} (1-x).
\end{equation}
This means that universality in $Z \rightarrow l \overline {l}$ would still
hold if $x=1$.  [In our previous papers, it was assumed that $v_{01}^2+v_{02}^2
<< v_{03}^2$, i.e. $x<<1$, in which case $\Gamma_\tau < \Gamma_{e,\mu}$ would
be required.]  As for $sin^2\theta_l$, if we take the average over
$l=e,\mu,\tau$, then
\begin{equation}
sin^2\theta_{eff} \simeq s_0^2 \left[ 1- \left( {1-s_0^2} \over {1-2s_0^2}
\right) \rho_{rad} + (\xi-1) \left[ {{1-x} \over 3} + { {y(1-s_0^2x)} \over
{1-2s_0^2} } \right] \right],
\end{equation}
where $s_0^2(1-s_0^2) \equiv \pi\alpha(M_Z^2)/\sqrt2G_FM_Z^2$.  The
corresponding average value of $\rho_l$ is
\begin{equation}
\rho_{eff}~\simeq~1+\rho_{rad}-(\xi-1) \left[ {2 \over 3} (1-x) + xy \right].
\end{equation}

\section{Comparison with Data}

	As mentioned already, the $\tau$-lifetime discrepancy implies that
$\xi-1=0.015 \pm 0.008$.  To pin down $x$ and $y$, we use Eqs. 6,7,8 and
compare with the present LEP data.  We also use $s^2 = 0.231 \pm 0.006$
(in the on-shell renormalization scheme) from neutrino data and compare what
it predicts for $M_Z$ as a function of $m_t$ in the standard model to the
observed value $M_Z = 91.175 \pm 0.021~GeV$, which gives us another constraint.
The combined 1989 and 1990 data of all 4 LEP collaborations are now published.
\cite{3}  The preliminary results of the 1991 run have also recently become
available.\cite{4}  We put these together and find $\rho_{eff} = 0.9990 \pm
0.0032$, $sin^2\theta_{eff} = 0.2322 \pm 0.0015$, $\Gamma_{e,\mu} = 83.44
\pm 0.27~MeV$, $\Gamma_\tau = 83.38 \pm 0.60~MeV$, and $s_0^2 = 0.2338 \pm
0.0005$, where $\alpha^{-1}(M_Z^2) = 127.9 \pm 0.2$ has been used.  Taking
into account the $0.19~MeV$ reduction of $\Gamma_\tau$ due to $m_\tau$, we
obtain from Eq. 6 the following restriction on $x$:
\begin{equation}
1-{0.0030 \over {\xi-1}} < x < 1+{0.0045 \over {\xi-1}}.
\end{equation}
Consider now the standard-model limit: $\xi-1=0$.  Assuming that $\rho_{rad}$
is dominated by $m_t$, we then find from $\rho_{eff} < 1.0022$ that $m_t <
84~GeV$, which is already ruled out by the CDF result $m_t > 91~GeV$.
However, if we allow $\xi-1 = 0.015 \pm 0.008$ as present data indicate,
then this particular restriction on $m_t$ is removed.  On the other hand,
the constraint from neutrino data which gives $m_t < 147~GeV$ for $\xi-1=0$
is only extended a little to $m_t < 159~GeV$, the maximum occurring at
$\xi-1=0.007$, $x=0.57$, and $y=0.91$.  Hence this model would still require
$m_t$ to be small enough to be experimentally accessible in the near future.

	In Fig. 1 we show the allowed region in $y$ versus $x$ for $\xi-1=0.01$
and $m_t=150~GeV$.  It is seen that $y$ is bounded from below by $\rho_{eff}$
and the $\nu N$ data, and from above by $sin^2\theta_{eff}$.  If $\xi-1$ is
increased to 0.02, then $\rho_{eff}$ provides both upper and lower bounds on
$y$, and if $m_t$ is decreased to 120 $GeV$, the $\nu N$ data do not restrict
$y$ at all, as shown in Fig. 2.  Recall that $0<y<1$ is required by definition
and $x$ is restricted by $\xi-1$ according to Eq. 9.

\section{Future Implications}

	The effects due to nonuniversality are naturally very small at
present energies.  They will probably never be a decisive test of our model.
However, we also predict the existence of a second set of $W$ and $Z$ bosons
with nearly degenerate masses approximately given by
\begin{equation}
M_{W'}^2~=~M_{Z'}^2~=~{{M_Z^2 (1-s_0^2)} \over {(\xi-1) (1-y) x}}.
\end{equation}
To maximize the denominator, we take $\xi-1=0.023$, $y=0$, $x=0.974$, and
$m_t=91~GeV$, whereby we find a lower bound of $533~GeV$ for $M_{W'}=M_{Z'}$.
There is no upper bound at present, because the $y=1$ limit cannot be ruled
out by the data.  However, once $m_t$ is measured, this may become possible.
In any case, we expect it to be no more than a few $TeV$.  Hence these new
vector gauge bosons are predicted to be discovered at future accelerators
such as the SSC or LHC.  Their decay rates into the third versus the first or
second generations are predicted to be in the ratio $(1-y)^2$ to $y^2$, and
they only have left-handed interactions.

\section{Conclusion}

	The idea that $e-\mu-\tau$ universality may be only accidental and
approximate as realized in our 1981 model is now being tested by precision
electroweak data.  The results are certainly not definitive, but they do
show the trends as predicted:  (1) the $\tau$ lifetime should be longer by
a small amount, and (2) the $\rho$ parameter measured at the $Z$ peak should
have an additional negative contribution of comparable magnitude to the
positive radiative correction proportional to $m_t^2$.  A future decisive
test of this model would be the discovery of new $W$ and $Z$ bosons with
nearly degenerate masses of a few $TeV$.

	This work was supported in part by the U. S. Department of Energy
under Contract No. DE-AT03-87ER40327.
\newpage
\bibliographystyle{unsrt}

\end{document}